\documentclass{lhep}       

\journal{LHEP}
\vol{xx}
\jyear{2023}
\pages{xxx} 

\received{xx March 2023}
\published{xx yyyy 2023}

\def\be{\begin{equation}}
\def\ee{\end{equation}}
\def\ba{\begin{eqnarray*}}
\def\ea{\end{eqnarray*}}

\newcommand{\prl}{Phys.\ Rev.\ Lett.\ }
\newcommand{\prd}{Phys.\ Rev.\ D }

\newcommand{\pl}{Phys.\ Lett.\ }

\newcommand{\myitem}[1]{#1~}

\newcommand{\cL}{{\cal L}}
\newcommand{\sign}{\mbox{sign}}

\newcommand{\degK}{^\circ\mathrm{K}~}
\newcommand{\power}[1]{\times 10^{#1}}
\newcommand{\LPl}{\Lambda_{\rm Pl}}

\newcommand{\MPl}{M_{\rm Pl}}
\newcommand{\mpl}{M_{\rm Pl}}
\newcommand{\tpl}{t_{\rm Pl}}

\newcommand{\gv}{\mbox{GeV}}

\newcommand{\ev}{\mbox{eV}}

\newcommand{\D}{\mathrm{d}}
\newcommand{\E}{\mathrm{e}}

\newcommand{\mbo}[1]{$#1$}
\newcommand{\semis}{\;;\;\;}

\newcommand{\MSb}{$\overline{\mathrm{MS}}$ }
\newcommand{\epo}{\,.}

\newcommand{\bit}{\begin{itemize}}
\newcommand{\eit}{\end{itemize}}

\newcommand{\SU}{\mathit{SU}}

\newcommand{\mz}{M_Z^2}
\newcommand{\mw}{M_W^2}
 
\renewcommand{\mbo}[1]{$#1$ }
\newcommand{\lpl}{\LPl}

\newcommand{\ha}{\frac12}
\newcommand{\mydot}[1]{{\raisebox{1.3ex}{$\cdot$}}\!\!\!{#1}}
\newcommand{\sinW}{\sin^2 \Theta_W}
\newcommand{\cosW}{\cos^2 \Theta_W}

\begin{document}

\title{Is the Higgs Boson the Master of the Universe?}

\author{Fred Jegerlehner\auno{1}\auno{2}}
\address{$^1$Deutsches Elektronen-Synchrotron DESY, Platanenallee 6, 15738 Zeuthen,\\
  Germany}
\address{$^2$Humboldt-Universit\"at zu Berlin, Institut f\"ur Physik,
       Newtonstrasse 15, 12489 Berlin,\\ Germany}

    \begin{abstract}
  The discovery of the Higgs particle has yielded a
  specific value for the mass of the Higgs boson, which, depending on
  some technical details in the calculation of the \MSb parameters
  (relevant for the high energy range) from the physical parameters
  (measured in low energy range), allows the Standard Model (SM) to
  hold up to the Planck scale about $\lpl \sim 10^{19}~\gv$. One then
  has the possibility that the Higgs boson not only provides mass for
  all SM-particles but very likely also has supplied dark energy that
  inflated the young universe shortly after the Big Bang. The SM Higgs
  boson is a natural candidate for the Inflaton, and the Higgs boson
  decays are able to reheat the universe after inflation.  I argue
  that the structures of the SM evolve naturally from a Planck cutoff
  medium (ether) and thus find their explanation. That the SM is an
  emergent structure is also strongly supported by Veltman's
  derivation of the SM from some general principles, which we can
  understand as the result of a low-energy expansion. I emphasize the
  role of the hierarchy problem and the problem of the cosmological
  constant as causal for the Higgs inflation scenario.  After the
  discovery of the Higgs boson at 125 GeV, and considering the absence
  of beyond the SM particles at the LHC, a new view on the SM of
  particle physics and its role in early cosmology has become
  indispensable. Very likely, the spectacular Higgs discovery turned
  out to have completed the SM in an unexpected way, revealing it as
  an inescapable emergence which shapes the early universe.
\end{abstract}
\onecolumn
\thispagestyle{empty}
\begin{flushright}
  DESY-23-057\\
  HU-EP-23/08\\
\end{flushright}

\vspace*{0.5cm}

\begin{center}
  {\normalsize
    
  {\Large \bf Is the Higgs Boson the Master of the Universe?}\\

\vspace*{0.5cm}

  {\large \bf Fred Jegerlehner}\\

\vspace*{0.5cm}

{Deutsches Elektronen-Synchrotron DESY, Platanenallee 6, 15738 Zeuthen,\\ Germany}\\
  
{Humboldt-Universit\"at zu Berlin, Institut f\"ur Physik,
       Newtonstrasse 15, 12489 Berlin,\\ Germany}\\

}

\vspace*{0.5cm}

  {\bf Abstract}\\

\end{center}

\vspace*{0.5cm}

\noindent
{\normalsize 
  The discovery of the Higgs particle has yielded a
  specific value for the mass of the Higgs boson, which, depending on
  some technical details in the calculation of the \MSb parameters
  (relevant for the high energy range) from the physical parameters
  (measured in low energy range), allows the Standard Model (SM) to
  hold up to the Planck scale about $\lpl \sim 10^{19}~\gv$. One then
  has the possibility that the Higgs boson not only provides mass for
  all SM-particles but very likely also has supplied dark energy that
  inflated the young universe shortly after the Big Bang. The SM Higgs
  boson is a natural candidate for the Inflaton, and the Higgs boson
  decays are able to reheat the universe after inflation.  I argue
  that the structures of the SM evolve naturally from a Planck cutoff
  medium (ether) and thus find their explanation. That the SM is an
  emergent structure is also strongly supported by Veltman's
  derivation of the SM from some general principles, which we can
  understand as the result of a low-energy expansion. I emphasize the
  role of the hierarchy problem and the problem of the cosmological
  constant as causal for the Higgs inflation scenario.  After the
  discovery of the Higgs boson at 125 GeV, and considering the absence
  of beyond the SM particles at the LHC, a new view on the SM of
  particle physics and its role in early cosmology has become
  indispensable. Very likely, the spectacular Higgs discovery turned
  out to have completed the SM in an unexpected way, revealing it as
  an inescapable emergence which shapes the early universe.
  
  }
    
\vspace*{0.5cm}

\vfill
\noindent\rule{8cm}{0.5pt}\\

\setcounter{page}{0}
\newpage
\twocolumn

\maketitle


\section{Introduction and Overview}
When the ATLAS and CMS collaborations~\cite{ATLAS,CMS} finally
discovered the Higgs boson~\cite{Englert:1964et,Higgs:1964ia} at 125
GeV at the LHC at CERN, I immediately got excited - not so much
because it turned out that the Higgs boson existed, of that, I had no
doubt, but because of its mass. The mass of the Higgs boson fell
almost perfectly within the narrow window that allows extrapolation of
the electroweak Standard Model (SM) up to the Planck scale. I first
learned of the existence of such a possibility from the study
\cite{Hambye:1996wb} about 16 years earlier, after the discovery of
the top quark at 172 GeV by the CDF collaboration at the Tevatron at
Fermilab. At that time, the Higgs boson mass was the only free SM
parameter. It was obvious that there must be a conspiracy at work here
between the heaviest SM particles. Suddenly, there seemed to be a
direct connection between the SM at accelerator energies and early
cosmology arising from a Planck-scale state. The Higgs boson was
required to save renormalizability and the predictive ability of the
electroweak part of the SM, and renormalizability now turned to be an
emergent property of the low energy effective tail of a Planck
medium~\cite{Jegerlehner:2013cta}.  The most important result of my
analysis: when going to higher and higher energy, well below the
Planck scale, there is a phase transition from the Higgs phase
(spontaneously broken case) to the symmetric state, where all
particles except the Higgs boson become massless. In the disordered
phase, there are four physical Higgs scalars, which are very heavy and
provide enormous Dark Energy (DE). Dark energy is equivalent with the
presence of a Cosmological Constant (CC). Dark energy triggers the
inflation of the early
universe~\cite{Guth:1980zm,Linde:1981mu}\footnote{ Close below the
Planck scale, we start to see the bare SM with the bare short-distance
effective parameters and, in particular, a very heavy Higgs boson,
which, at some stage, will be moving very slowly only, i.e., the
potential energy
$V(\phi)=\frac{m^2}{2}\phi^2+\frac{\lambda}{24}\phi^2$ then is large
the kinetic energy $\ha\, \mydot{\phi}^2$ is small. The Higgs boson
contributes to the energy-momentum tensor of Einstein's gravity
equation by providing a pressure $p=\ha\,\mydot{\phi}^2-V(\phi)$ and an
energy density $\rho=\ha \, \mydot{\phi}^2+V(\phi)$. Shortly after the
Big Bang, the slow--roll condition \mbo{\ha \, \mydot{\phi}^2\ll
  V(\phi)} is satisfied such that given $p\approx-V(\phi)\semis \rho
\approx +V(\phi)$ we find $p=-\rho$ the dark energy equation of
state. DE has been established by observation, through the Cosmic
Microwave Background
(CMB)~\cite{Mather1990,Smoot1992,Bennett:2012zja,Ade:2013uln} and
Super Nova (SN) counts~\cite{Riess:1998cb,Perlmutter:1998np}. In any
case, the SM Higgs boson in the early universe provides substantial
dark energy acting as anti-gravity that inflates the universe.}.
Inflation then tunes the universe to the critical density of flat
space, whereof the dark energy can only be a fraction, in agreement
with observation.

The Higgs boson discovery and the absence of beyond-the-SM physics at
experimentally accessible energy scales promotes a
``change of the game'' and requires a new paradigm. It also puts the
``cosmic bridge’', relating particle physics with events that must have
happened in the early universe, on a new basis.

While DE can be explained to be provided by the SM Higgs system, the
origin of Dark Matter (DM) remains a mystery. However, it could be
that DM is inherent in the Planck medium, similar to the displacement
field of the ion system in a superconductor. It is well known that we
may understand the main aspects of the BCS theory of superconductivity with a two-fluid
model, where the compressible electron gas and the related phonons
propagate in an incompressible ion background. Two types of collective
vibrational fields occur. The phonons populate the long-range branch
and the displacement field of the essentially incompressible ion
system, which could act as a DM field. This possibility, of course,
needs to be investigated first.

Detailed calculations based on the ``125 GeV Higgs boson completed
SM'' support to the greatest possible extent the points which I will
list in the following.  All statements result from calculations based
on the extrapolation of established SM physics (bottom-up approach).
For details I refer to my
papers~\cite{Jegerlehner:2013cta,Jegerlehner:2013nna,Jegerlehner:2014mua,Jegerlehner:2018zxm,Jegerlehner:2021vqz}
and references therein. My discussion relies on analyzing the
properties of the bare SM Higgs potential. The investigation has been
extended in~\cite{Jegerlehner:2018zxm} to the effective potential,
which includes the higher-order corrections. The analysis shows that
the main pattern, dictated by the Planck-scale amplified power
corrections, is barely affected by incorporating subleading
corrections.
\section{the SM as a low energy effective tail of a Planck cutoff medium}
The SM is extendable up to the Planck cutoff, and perturbative
calculations remain valid. At the Planck scale, where gravity unifies
with SM physics, a widely unknown Planck cutoff medium must
exist\footnote{The Planck length is a precisely known fundamental
parameter which must have its realization in a cutoff medium. A cutoff
far above accelerator energies $E$ calls for a Low Energy Expansion
(LEE) as $E/\LPl$ is a tiny parameter, and all effects $x^n$ for $n>0$
appear to be absent at accelerator energies.}, and the relationship
between renormalized and bare (cutoff dependent) parameters gain a
physical meaning which must be taken seriously. The SM at energies
sufficiently far below the Planck scale appears as a Low Energy
Effective Theory (LEET), specifically, a Quantum Field Theory
(QFT). That QFTs can be the result of a low energy expansion was discovered by Ken
Wilson in his avenue to solve the problem of critical phenomena of
condensed matter systems~\cite{Wilson:1971bg}. The renormalizable QFT structure, as
well as the non-Abelian gauge symmetries~\cite{Cornwall:1973tb},
realized within the SM, turn out to be emergent. These fundamental
properties show-up because we do not see much of the details of the
cutoff system when we are far away.

\section{Surprises concerning the SM\\ running parameters}
After the completion of the SM by the Higgs boson discovery, the
masses and the related couplings show amazingly deep conspiracy (all
heavy SM particles exhibit masses and related couplings in the same
ballpark\footnote{Generally claimed to be unnatural because of a
supposed Higgs hierarchy problem.}), which in particular reveals that
the SM vacuum likely remains stable up to the Planck
scale\footnote{The experts in this field find a metastable
vacuum~\cite{Yukawa:3,Degrassi:2012ry,Buttazzo:2013uya,Bednyakov:2015sca,Kniehl:2015nwa},
while they find vacuum stability to be missed by 1.5 $\sigma$. In view
that some of the SM parameters seem to report underestimated
uncertainties, there is sufficient space for vacuum stability to turn
out true in the future.}. Not only this but all couplings except from
the hypercharge $U(1)_Y$ coupling $g_1=g'$ (which only increases very
moderately) turn out to behave asymptotically free~\cite{AF}. Not only
the gauge couplings $g_2=g$ and $g_3=g_s$, related to the weak
$\SU(2)$ and the strong $\SU(3)_c$ non-Abelian gauge groups,
respectively, but also the top-quark Yukawa-coupling $y_t$ and the
Higgs boson self-coupling $\lambda$ (both intrinsically infrared
free), within the SM, by a ``collaboration'' between the leading SM
couplings, get transmuted to behave as asymptotically free (i.e.,
ultraviolet free), decreasing with increasing energy-scale, and
improving the convergence of the perturbation expansion. The change
from screening (IR free) to anti-screening (UV free) results because
the Higgs mass, top-quark mass, and gauge boson masses conspire
appropriately\footnote{More details are provided in Sect.~4
in~\cite{Jegerlehner:2013cta}}.

This conspiracy is seen already in the one-loop approximation in the
gaugeless limit (\mbo{g_1,g_2=0}): the top-quark Yukawa-coupling to be
anti-screening requires QCD effects to dominate
$g_3>\frac{3}{4}\,y_t$, what is the case; the Higgs self-coupling to
be anti-screening requires top-quark Yukawa-coupling effects to
dominate $y_t^2 > \frac{2}{3\,(\sqrt{5}-1)}\lambda$, what is satisfied
as well. We further require the Higgs potential
$$V=\frac{m^2}{2}\,\phi^2+\frac{\lambda}{24} \phi^4$$
to satisfy vacuum stability $\lambda(\mu) >0$ up to $\mpl$.

\begin{figure}[h!]
\centering
\includegraphics[height=1.6in]{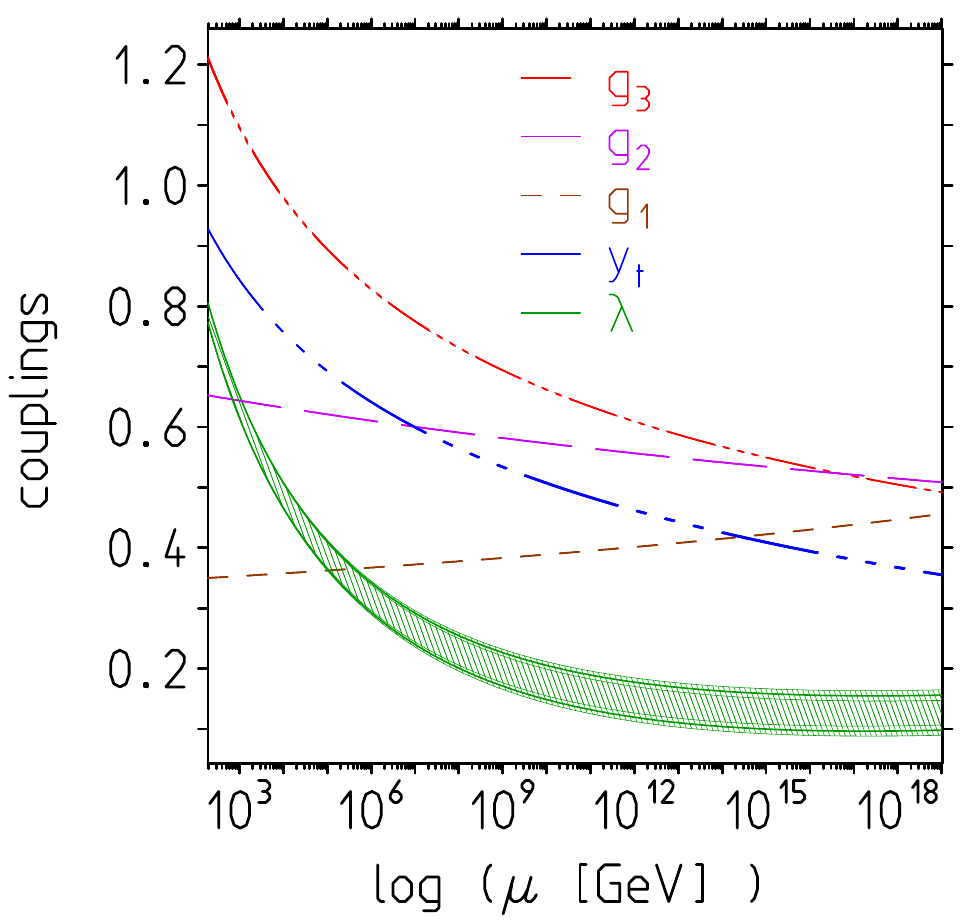}
\includegraphics[height=1.6in]{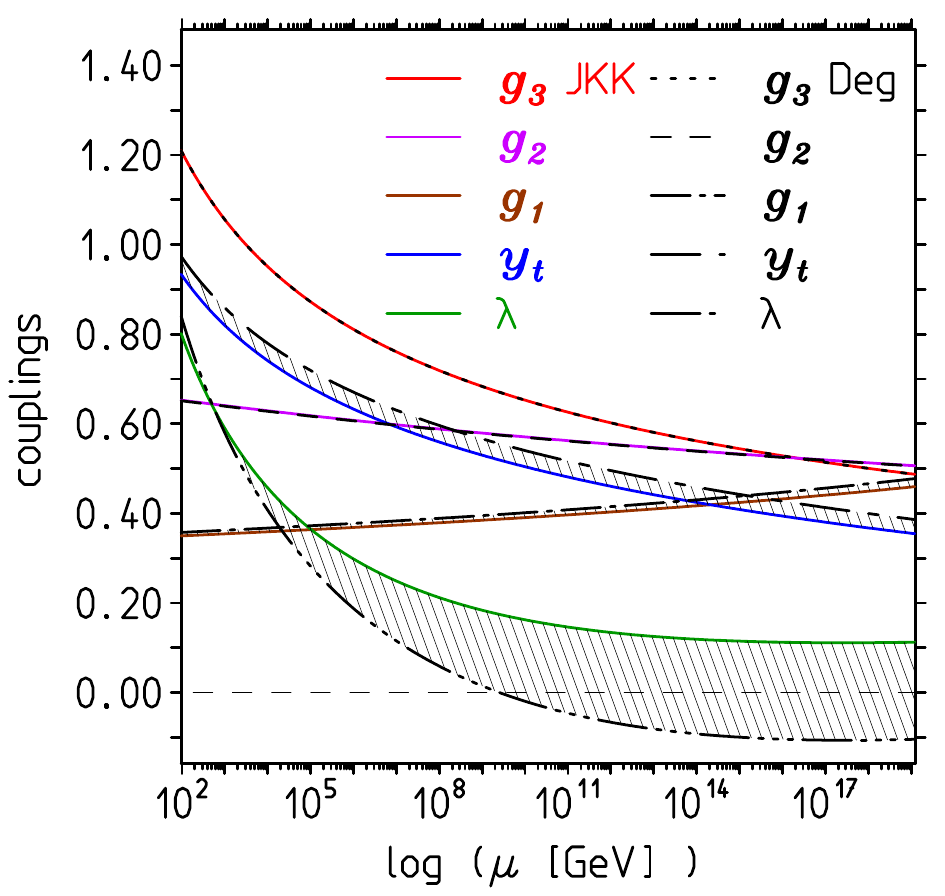}
\caption{Left: The SM dimensionless couplings in the \MSb scheme as a
  function of the log of the renormalization scale for $
  M_H=124-126~\gv$ (green band)~\cite{Jegerlehner:2013cta}. Input
  parameters as given in the text. Right: Results depend on
  implementations of the two-loop on-shell vs. \MSb parameter
  matching. {\tt JKK} labels the results
  from~\cite{Yukawa:3,Jegerlehner:2013cta} and shows a stable
  vacuum. {\tt Deg} labels the results from~\cite{Degrassi:2012ry}, which
  find a breakdown of vacuum stability far below $\LPl$. Including
  higher-order corrections to the effective Higgs potential turns the
  instability into metastability (see Fig.~\ref{fig:VeffLogcomp} below).}
\label{fig:runcoup}
\end{figure}

Whether the latter condition holds depends on very precise input
parameters and precise matching conditions between the physical low
energy parameters and the \MSb parameters, which are the appropriate
parameters for the extrapolation to very high energies\footnote{Our
\MSb renormalization scheme is gauge invariant as it includes the
tadpole diagram contributions (see
~\cite{Fleischer:1980ub,Jegerlehner:2012kn}), which commonly are
omitted. In our context where bare parameters in the high energy phase
are physical, the proper parametrization is mandatory because only the
gauge invariant \MSb parameters are in direct correspondence with the
bare parameters upon the identification of the \MSb renormalization
scale $\mu$ with the cutoff $\Lambda$.}. My \MSb input\footnote{I
updated the parameters according to the shifts of $M_t$ from 173.50
GeV to 172.76 GeV and $M_H$ from 126.0 GeV to 125.2 GeV.} at scale
$\mu=M_Z$ is~\cite{Jegerlehner:2013cta} $g_3=1.2200$, $g_2=0.6530$,
$g_1=0.3497$, $y_t=0.9307$ and $\lambda=0.7968$. Given this set of
start-up parameters, we display the solutions of the SM
Renormalization Group (RG) equations\footnote{The RG equations $\mu
\frac{d}{d \mu}\: g_i(\mu)=\beta_i(g_i)\;(i=1,2,3)$, $\mu \frac{d}{d
  \mu} y_t(\mu) =\beta_{y_t} (y_t,\cdots)$ and $\mu \frac{d}{d \mu}
\lambda(\mu) =\beta_\lambda (\lambda,\cdots)$ for the \MSb couplings
$g_i,y_t$ and $\lambda$ define the \MSb running couplings as functions
of the \MSb energy scale $\mu$. The \MSb renormalization scheme is the
appropriate parametrization for the high energy behavior $\mu \gg M_t$
where $M_t$ is the top-quark mass, the largest of the SM masses.  The
RG coefficient-functions $\beta_i$ are known to
3-loops~\cite{Mihaila:2012fm,Bednyakov:2012rb,Chetyrkin:2012rz}, and
some of them beyond.}  in Fig.~\ref{fig:runcoup} [left panel] together
with results obtained by other groups [right panel].
 The plots reveal a high sensitivity (shaded range) on the input
 parameters. While the running of the Higgs boson self-coupling
 $\lambda$ is highly dependent on the value of $y_t$ at $M_Z$, the
 other parameters are in good agreement. At \mbo{\mpl} I get
 $g_3=0.4886$, $g_2=g=0.5068$, $g_1=g'=0.4589$, $y_t=0.3510$ and
 $\lambda=0.1405$. Note that, according to the mass-coupling relations, it is the root $\sqrt{\lambda}=0.3748$,
 which is comparable with the other couplings.

This surprising pattern of the leading SM couplings validates the SM
to work to the highest energy and heavily relies upon Higgs masses
constrained to a narrow window in parameter space!

\section{The issue of quadratic divergences and Veltman's Infrared - Ultraviolet Connection}
\label{sec:jump}
\begin{figure}
\centering
\includegraphics[height=1.6in]{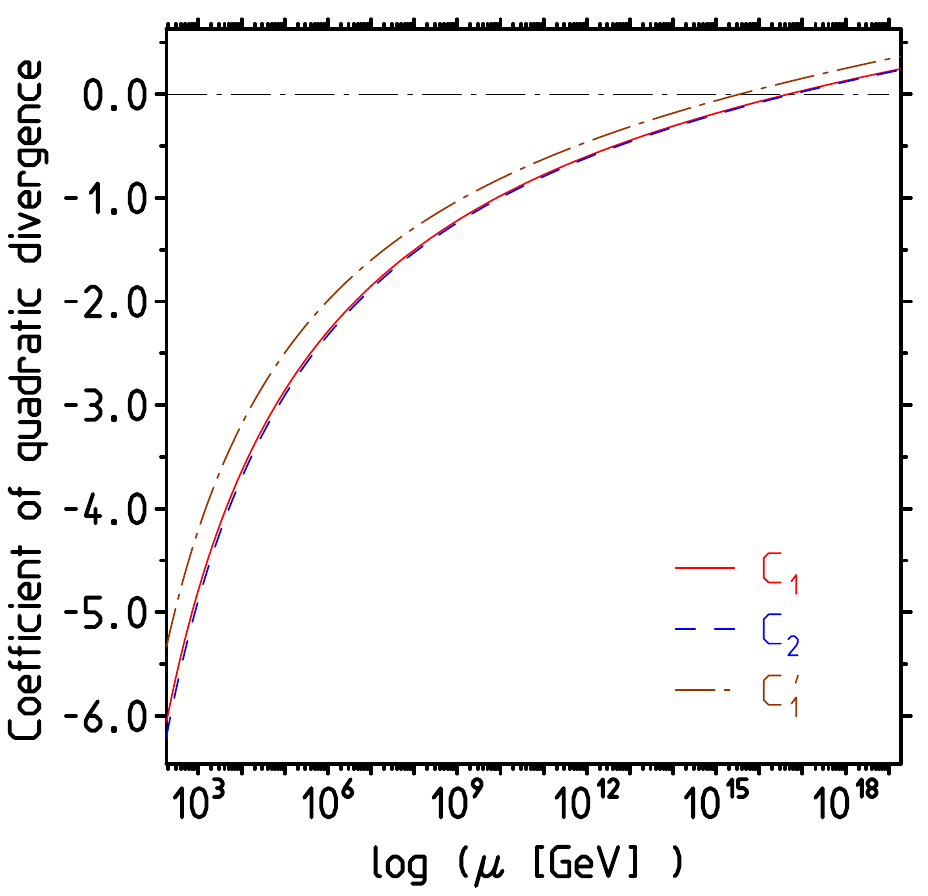}
\includegraphics[height=1.6in]{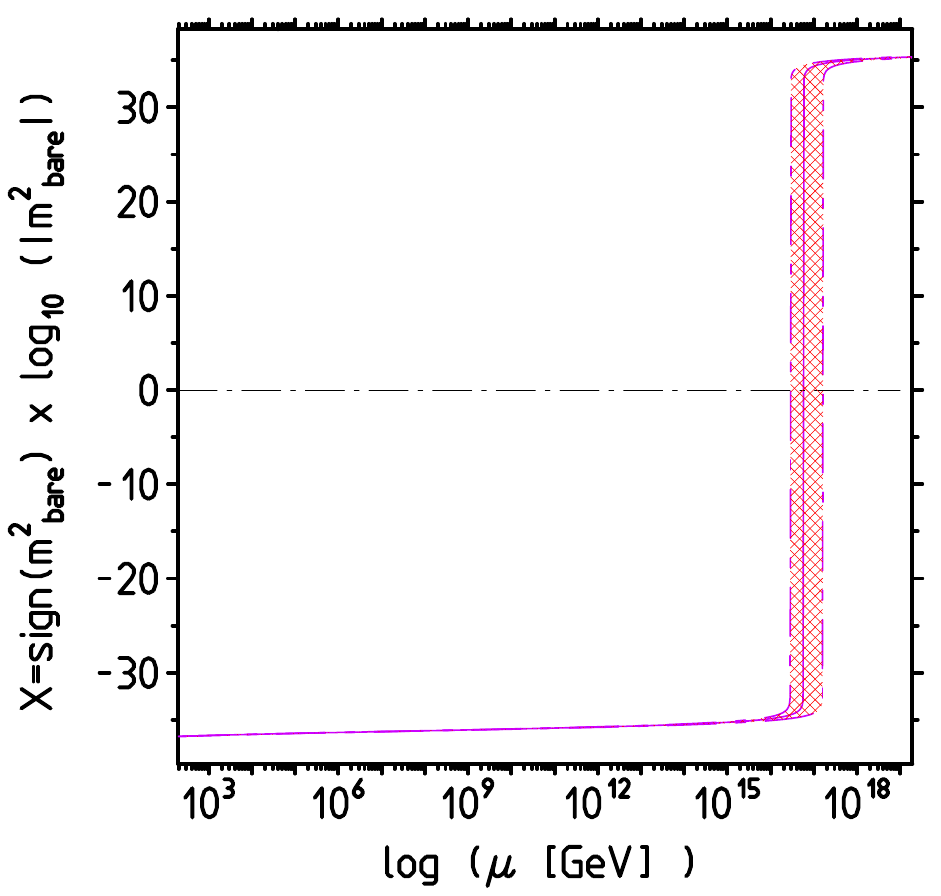}
\caption{The phase transition in the SM. Left: the zero in \mbo{C_1}
  (1-loop) and \mbo{C_2} (2-loop) for \mbo{M_H=125.2\pm
    0.2~\gv}. Right: shown is $X=\sign(m^2_{\rm bare})\times \log_{10}
  (|m^2_{\rm bare}|)$, which represents $m^2_{\rm bare}=\sign(m^2_{\rm
    bare}) \times 10^{X}$. The band covers the range $M_H=124-126~\gv$. The difference between $C_1$
  (1-loop)~\cite{Veltman:1980mj} and $C_2$
  (2-loop)~\cite{Hamada:2012bp,Jones:2013aua} results is
  marginal. $C_1'$ results after Wick ordering the field monomials of
  the potential. Note that
  ${m'}^{2}=C_1'\,\Xi=m^2+\frac{\lambda}{2}\,\Xi$ is the proper effective
  physical mass.}
\label{fig:jump}
\end{figure}

The expansion of the universe implies an energy scan in time, from
Planck temperatures $T_{\rm Pl} \sim 1.4\power{32}~\degK$ down to
$T_{\rm CMB} \sim 3^\circ$K CMB of
today. Since all SM couplings are energy dependent, as ruled by the
RG, radiative corrections of the Higgs boson mass
are affected in a particular manner. It is a fact that the Higgs boson
mass plays a special role because it is the only mass not protected by
gauge or chiral symmetry in the ultraviolet (UV) region. Also, the
Higgs field is special, as it is the only field that directly couples
to the vacuum and to all massive particles, i.e., to all particles
except for the photon and the gluons. Mass generation via the Brout-Englert-Higgs
mechanism results from the spontaneous breakdown of the $\SU(2)_L$
weak gauge symmetry and chiral symmetry. Spontaneous symmetry
breaking is a low-energy phenomenon (long-range order, Low Energy (LE) phase), and
when going to higher energies, at some point, the long-range order
gets destroyed while the symmetry gets restored. At the Planck scale,
we expect the SM to reside in the symmetric High Energy (HE) phase. Here Veltman's
``The Infrared - Ultraviolet Connection'' comes into play which
concerns the renormalization of the Higgs boson
mass~\cite{Veltman:1980mj} (also
see~\cite{Hamada:2012bp,Jones:2013aua}). The related counterterm we
may write as\footnote{For the physical masses in the LE phase, we use
capital letters like $M_H$, and for \MSb masses lowercase letters like
$m_{H0}$. The \MSb masses agree with the bare masses. For the Higgs
potential mass in the HE phase, we usually omit the label “bare”: $m^2=m^2_{\rm bare}$.
Note that $m^2= m^2_{H0}/2$.}
$$\delta m^2_H=m^2_{H0}-M^2_H=C_1\,\Xi\,;\: C_1={2\,\lambda+{\small
    3/2}\, {g'}^{2}+{\small 9/2}\,g^2}{-12\,y_t^2}$$ with $\Xi=
\frac{\LPl^2}{16 \pi^2}$ the quadratic Planck scale enhancement factor
and $C_1$ the one-loop coefficient function after neglecting small
lighter fermion contributions. By $m^2_{H0}$ we denote the bare mass,
by $M^2_H$ the renormalized one. Taking into account the scale
dependence of the SM couplings, the coefficient \mbo{C_1(\mu)}
exhibits a zero, for $M_H=125~\gv$ at about $\mu_0\sim 7
\power{16}~\gv$, clearly but not far below $\mu=M_{\rm
  Planck}$~\cite{Jegerlehner:2013cta}, i.e., there exists a point
where renormalized and bare mass coincide. Below the zero, the Higgs
mass counterterm changes sign and becomes large negative\footnote{When
one continues to adopt $\LPl$ as a cutoff, for which, in the
renormalizable LEET, there is no reason. In the renormalizable QFT, a
cutoff is an auxiliary tool setting a renormalization scale chosen to
be large relative to the heaviest particle in the spectrum. As
physical quantities are RG invariant, the renormalization scale, after
all, may be chosen as appropriate.}, which triggers a vacuum
rearrangement, a phase transition (PT), see Fig.~\ref{fig:jump}, from
the symmetric to the spontaneously broken phase. Obviously, the
Brout-Englert-Higgs mechanism~\cite{Englert:1964et,Higgs:1964ia} is
triggered by the cooling down of the expanding universe, which
provides an energy scan where at some point the phase transition must
happen. Below the zero, the memory to the cutoff gets lost since now
we are in the renormalizable SM, and the latter is devoid of
observable cutoff effects.
The point is that the mass-coupling relations, that characterize the
Higgs phase, at the PT point exhibit a jump \\[2mm]
\begin{tabular}{lcllc}
\hline\noalign{\smallskip}
\mbo{M_W^2}&=&\mbo{\frac14\,g^2\,v^2\semis} & $\to$ & 0, \\
\mbo{M_Z^2}&=&\mbo{\frac14\,(g^2+g'^2)\,v^2\semis} & $\to$ & 0, \\
\mbo{m_f^2}&=&\mbo{\frac12\,y^2_f\,v^2\semis} & $\to$ & 0, \\
\mbo{M_H^2}&=&\mbo{\frac13\,\lambda\,v^2} & $\to$ & $m^2_{H0}\approx \frac{\LPl^2}{16\pi^2}\,C_1\,,$\\[2mm]
\hline
\end{tabular}

\smallskip
\noindent
and this discontinuity uncovers the special role of the Higgs boson
mass, which is that in the symmetric phase, indeed it gets enhanced
through the Planck cutoff.  In the symmetric phase, the bare Higgs
boson mass represents the effective physical mass. Also, note that the
Planck-scale enhancement refers to the symmetric phase only. In the
broken phase, $m^2_{\rm bare}$ formally yields a large negative
mass-square contribution, which never can describe a particle in the
physical spectrum. But it triggers a new non-trivial minimum of the
potential where all particles, including the Higgs state itself, get a
mass proportional to the Higgs VEV. We point out that the often
mentioned expectation that the Higgs VEV should be of Planck-scale
order is untenable. Remember that we have $v\equiv 0$ in the HE phase,
for which reason should we expect that in the LE phase $v \propto
\LPl$?

In this scenario, above the PT point at $\mu_0$, the SM is in the
symmetric phase, and we obtain the Higgs potential mass
\mbo{m^2=(m_H^2+\delta m_H^2)/2} given by (with \MSb parameters listed
earlier)
$$m^2\sim \delta m^2 \simeq
\frac12\,C(\mu=\mpl)\,\Xi \simeq \left(0.0295\,\mpl
\right)^2\,,$$
i.e., $m^2(\mpl)/\mpl^2\approx 0.87\power{-3}\epo$
Since the dominant counter-term is induced via radiation-correction
effects, the quadratic Planck-mass dependence turns out to be
substantially reduced. Actually, the corresponding enhancement of the
Higgs potential alone does not provide sufficient dark
energy to explain
the required amount of inflation~\cite{Jegerlehner:2014mua}.

\begin{figure}[h!]
\centering
\includegraphics[height=4cm]{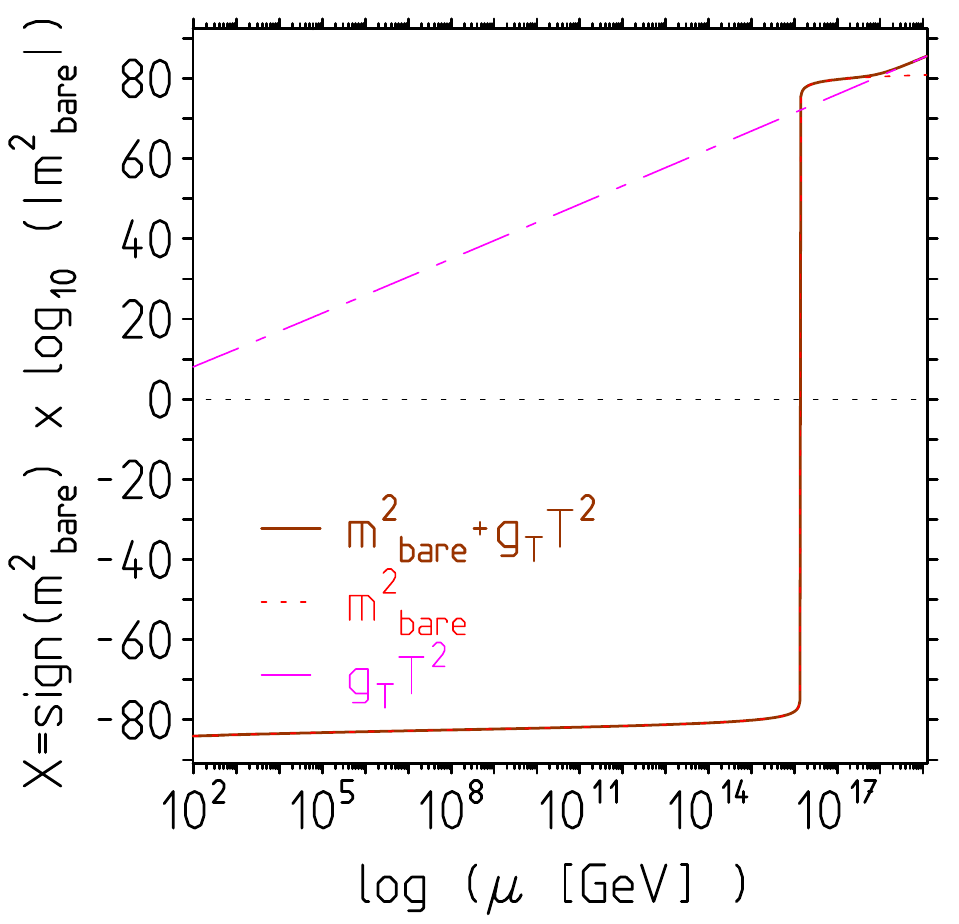}
\includegraphics[height=4cm]{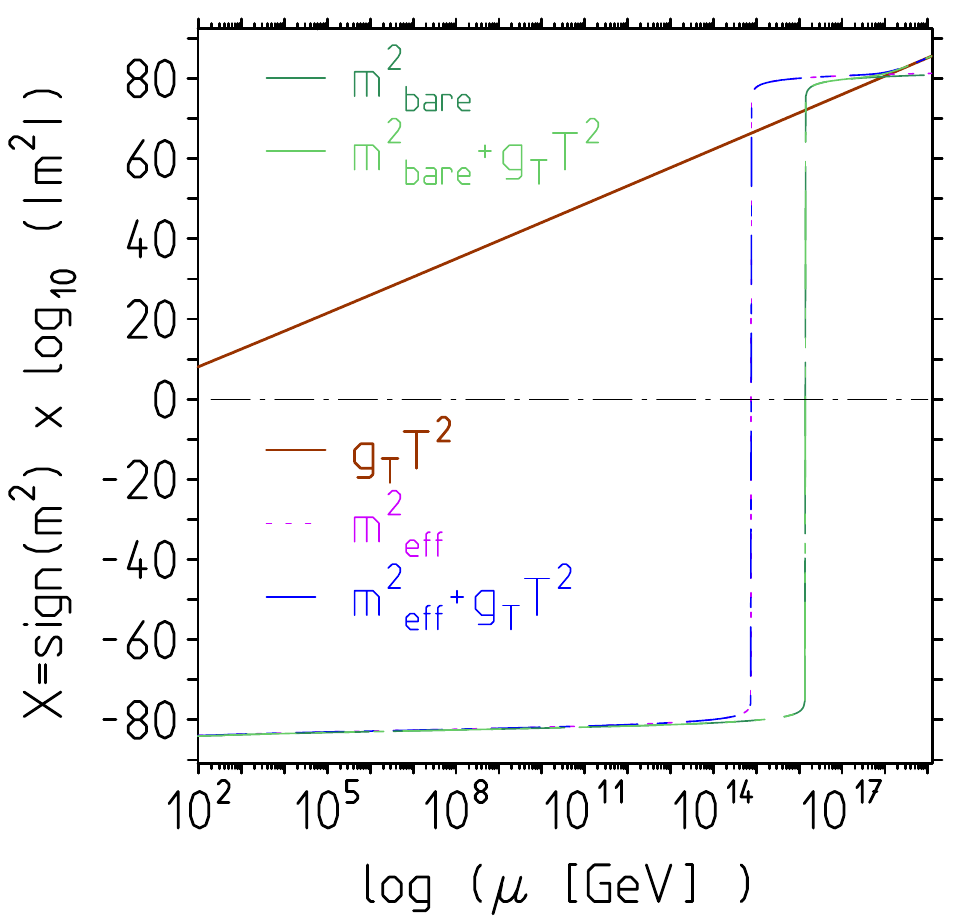}
\caption{$X$ as displayed in the right panel of Fig.~\ref{fig:jump}
including leading finite temperature correction to the potential
$V(\phi,T)=\ha\,(g_T\,T^2+m_0^2)\,\phi^2+
\frac{\lambda}{24}\,\phi^4+\cdots$ with $g_T=\frac{1}{16}
\left[3g_2^2+g_1^2+4y_t^2+\frac23\,\lambda\right]$ from~\cite{Dine:1992wr} affecting the phase transition
point. Left: for the bare case [\mbo{m^2,C_1}]. Right: with adjusted
effective mass from vacuum rearrangement
[\mbo{{m'}^2,C'_1=C_1+\lambda/2}]. In the case
\mbo{\mu_0} sufficiently below \mbo{\mpl}, the case displayed here,
finite temperature effects affect the position of the phase transition
little, while the change of the effective mass by the vacuum
rearrangement is more efficient. The finite temperature effect with
our parameters is barely visible.}
\label{fig:FT}
\end{figure}

We also have the effects that by Wick ordering the field monomials of
the potential, a shift of the effective mass to
${m'}^{2}=m^2+\frac{\lambda}{2}\,\Xi$ takes place. Using our values of
the \MSb input parameters, we obtain a shift of the PT
point\\ \centerline{\mbo{\mu_0 \approx 1.4 \power{16}~\gv \to
    \mu'_0\approx 7.7 \power {14}~\gv\,.}}  This shift implies that
the PT happens at a somewhat lower energy scale, i.e., later in the
history of the universe.  In the early hot universe, one also has to
take into account the finite temperature effects. However, because the
Planck-cutoff effects generate a large effective Higgs-boson mass, the
temperature effects are subdominant and thus barely change the PT
point, as Fig.~\ref{fig:FT} shows. Since finite temperature effects
increase proportional to $T^2$, these would become important when the
PT point $\mu_0$ would be closer to $\MPl$.

\section{Vacuum stability, the cosmological constant, and Higgs inflation}
\begin{figure*}[h!]
\centering
\includegraphics[height=1.6in]{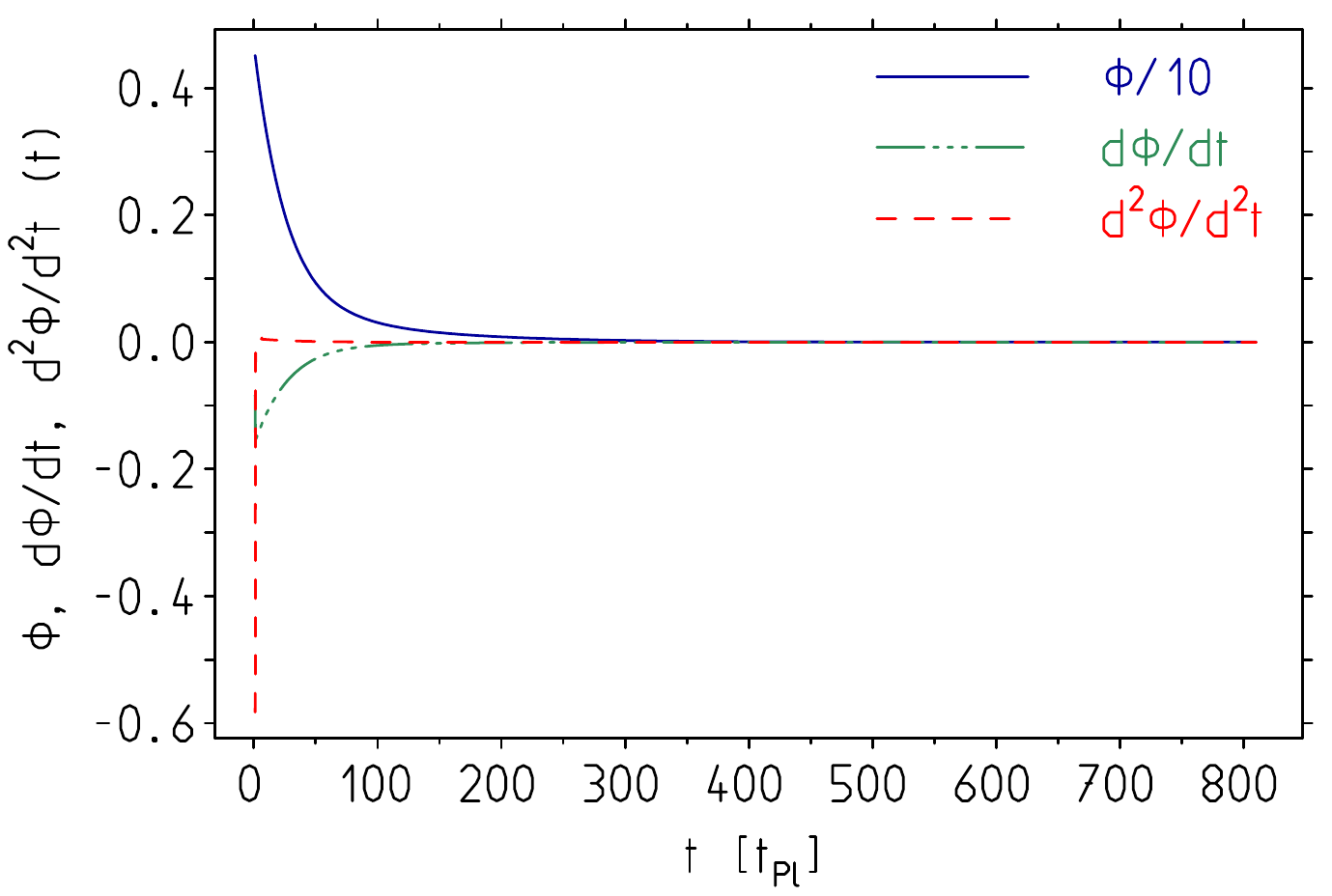}
\includegraphics[height=1.6in]{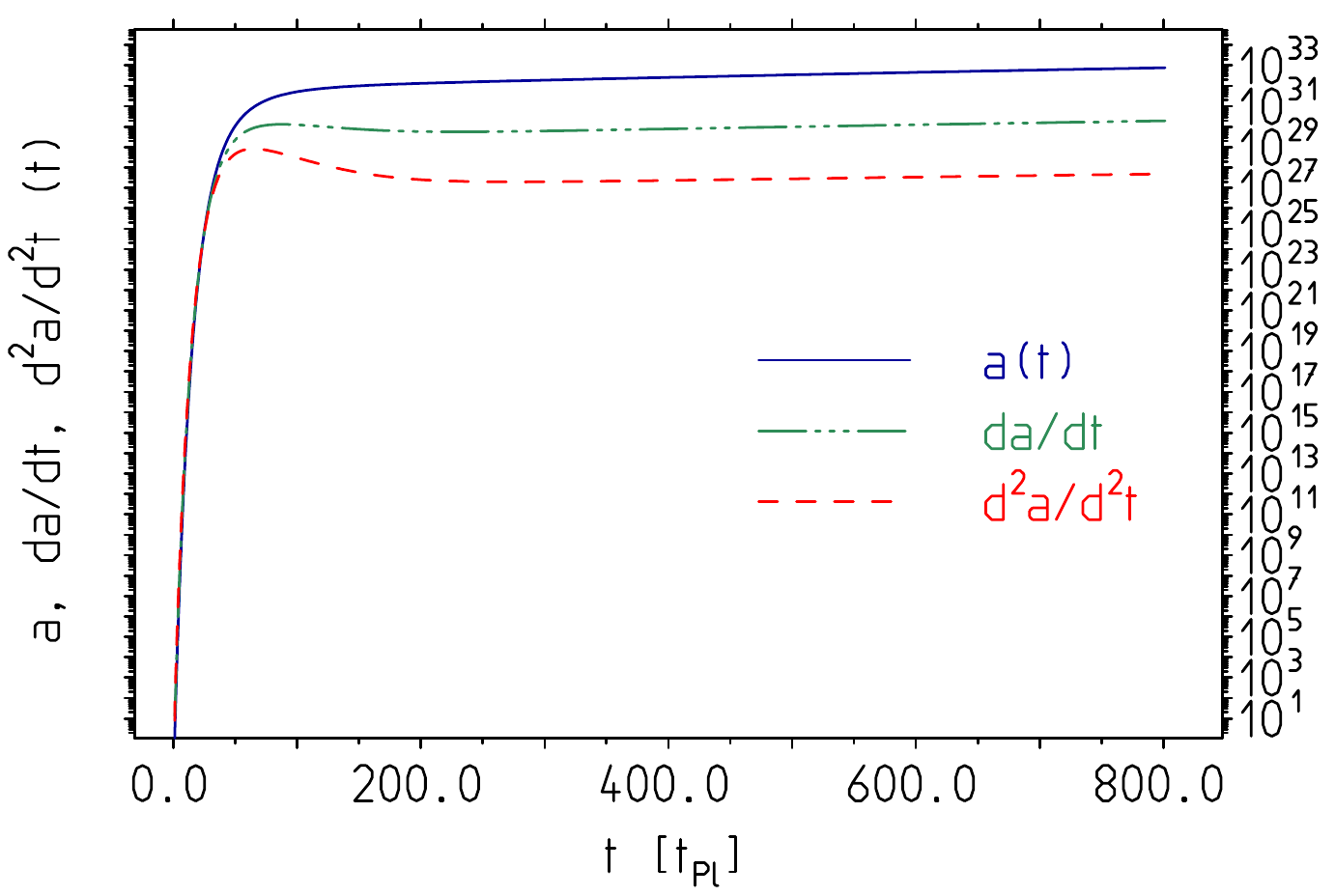}
\caption{The trans-Planckian Higgs field at $\tpl$ decays very fast (
$\phi(t)=\phi_0\,\E^{-E_0\,(t-t_0)}\,;\, 
E_0=\frac{\sqrt{2\lambda}}{3\sqrt{3}\ell}\approx4.3\power{17}~\gv;\,
V_{\rm int} \gg V_{\rm mass}\,,$
and $\phi(t)\approx \phi_0\,\E^{-E_0\,(t-t_0)}\,;\, 
E_0\approx\frac{m^2}{3\ell\,\sqrt{V(0)}}\approx6.6\power{17}~\gv;\,
V_{\rm mass} \gg V_{\rm int}$)
and inflation gets stopped soon. Left: the decaying Higgs field $\phi(t)$. Right:
the inflating radius of the universe $a(t)$.}
\label{fig:field}
\end{figure*}
\begin{figure*}[h!]
\centering
\includegraphics[height=1.6in]{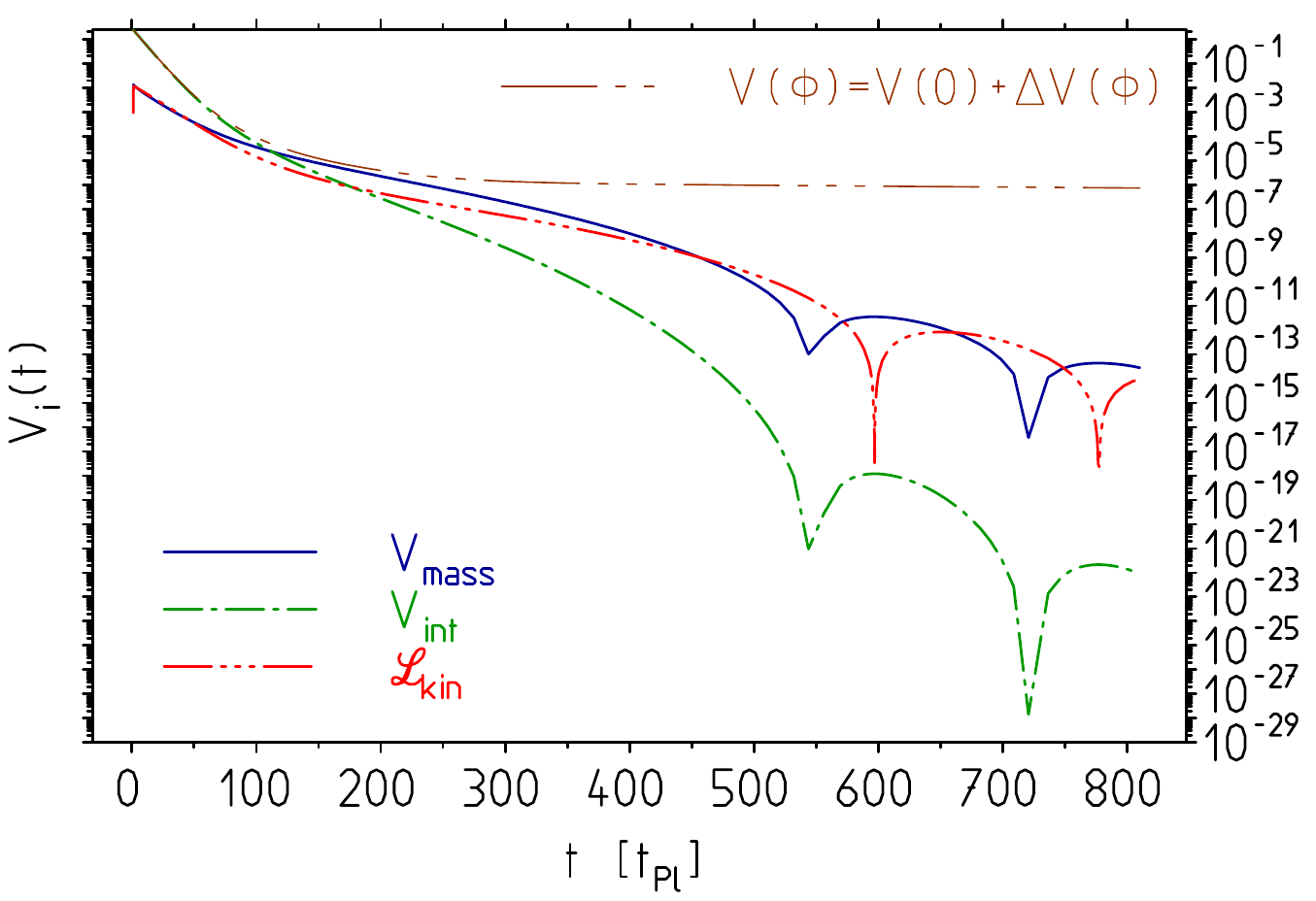}
\includegraphics[height=1.6in]{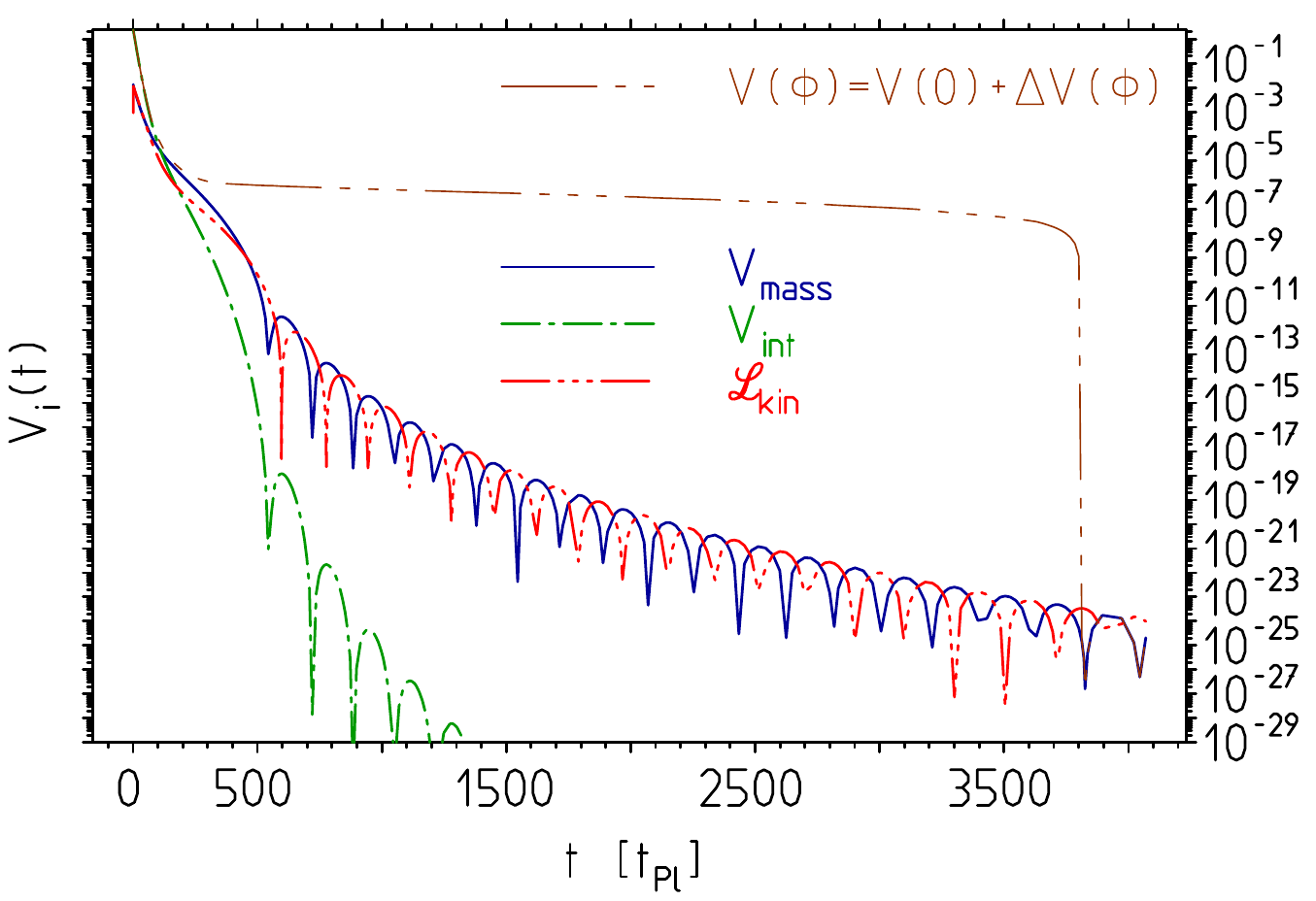}
\caption{SM dark energy contributions during inflation (left panel)
  and continued until the Higgs phase transition (right panel).
  Displayed are the mass-, interaction- and kinetic-term of the bare
  Lagrangian in units of $\MPl^4$ as a function of time.
  The evolution until symmetry breakdown and vanishing of the CC.
  Slow-roll inflation stops at about $t\approx
  450\,\tpl$ when $\cL_{\rm kin} \sim V_{\rm mass}\gg V_{\rm int}\,,$
  after which the Higgs field behaves as a damped quasi-free oscillating quantity.}
\label{fig:Lagrangian} 
\end{figure*}
As mentioned before, in the symmetric HE phase, the bare parameters
have a physical meaning. Here the SM directly attaches to the
Planck medium, and the Planck mass sets the relevant
scale. Furthermore, all four fields of the SM Higgs doublet field
$\Phi(x)$ represent heavy physical scalars of equal mass. Only one of the four Higgs
fields, we denote it $\phi$, is a neutral scalar and can couple to the
vacuum. $\phi$ is the doublet entry corresponding to the physical
Higgs field $H$ in the Higgs phase. While in the symmetric phase, we
have a vanishing Higgs VEV, i.e., $\langle \Phi(x)\rangle \equiv 0$,
expectation values of the singlet $\Phi^+\Phi$, and powers thereof,
are non-vanishing. So one obtains $\langle |\Phi(x)|^2\rangle
=\frac12\,\langle \phi^2\rangle \doteq \frac12\, \Xi$.  This implies
that the vacuum expectation value $$V(0)=\langle V(\phi) \rangle
=\frac{m^2}{2}\,\Xi+\frac{\lambda}{8}\,\Xi^2$$
of the potential is enhanced proportionally to $\LPl^4$. Obviously,
\mbo{\lambda} must be positive for the SM Higgs vacuum to be
stable. Besides an enhanced potential mass, we also get a large VEV of
the potential such that $V(\phi)= \Delta V(\phi)+V(0)$ is further
enhanced. $\Delta V(\phi)$ is the part depending explicitly on the
field $\phi$.
The effective CC counterterm has
a zero, which again is a point where renormalized and bare quantities
are in agreement:
$$\small \rho_{\Lambda\,{\rm bare}}=\rho_{\Lambda\,{\rm ren}}
+X(\mu)\:\Xi^2 \semis X(\mu)\simeq {5\,\lambda+3\,
  {g'}^{2}+9\,g^2}{-24\,y_t^2 }$$ with $X(\mu)=2C_1(\mu)+\lambda(\mu)$
exhibiting a zero close to the zero of $C_1(\mu)$, when
$2\,C_1(\mu)=-\lambda(\mu)$, with \mbo{\lambda(\mu)} small, which
happens at \mbo{\mu_{\rm CC} \approx 3.1 \power{15}~\gv}, between
\mbo{\mu_0 \approx 1.4 \power{16}~\gv} and \mbo{\mu'_0\approx 7.7
  \power {14}~\gv\,.}  It means that neither the hierarchy problem nor
the cosmological constant problem exists in the Higgs phase of the SM,
being the low-energy relict world we experience today.

A huge DE is needed to trigger inflation. DE transforms the otherwise
very unstable flat universe into a stable one, i.e., flat space turns
into an attractor. Since inflation tunes the total energy density to
be the critical density $\rho_{\rm crit}=\mu^4_{\rm crit}$ with
$\mu_{\rm crit}=0.00247~\ev$, of the flat space, DE can only be a
fraction of that, which we know is the case. CMB data tell us that
today we have $\rho_{\Lambda}=\mu^4_{0,\Lambda}$ with
$\mu_{0,\Lambda}=0.00171~\ev$, i.e., dark energy amounts 69\% of the
total density, and is the dominating part.

Since $V(0)> 0$ is \underline{positive} and quartically Planck-scale
enhanced, we obtain a sufficient amount of dark energy to get the
required amount of inflation, provided we assume\footnote{When DE
dominates the total energy density, the Friedmann equation
\mbo{\frac{\D a}{a}=H(t)\,\D t} predicts \mbo{a(t)=\exp Ht} an
exponential growth of the radius $a(t)$ of the universe. \mbo{H(t)} is
the Hubble constant \mbo{H\approx \ell\, \sqrt{V(\phi)}} where
$\ell^2=8\pi\,G/3$, G the Newton constant of gravity. The field
equation \mbo{\ddot{\phi}+3H\,\mydot{\phi}\simeq-V'(\phi)\,} implies
that inflation stops quite quickly as the field decays exponentially
(see Fig.~\ref{fig:field}). The fast decay of the field also implies
that the mass term dominates \mbo{V(\phi)\approx
  \frac{m^2}{2}\,\phi^2} before inflation ends. We then have a
harmonic oscillator with friction which implies Gaussian
inflation. The time dependence of the different terms of the
Higgs-Lagrangian is shown in Fig.~\ref{fig:Lagrangian}. It is
interesting to see which term dominates in which time period as time
evolves. Inflation flattens the universe as the curvature term
$k/a^2(t)\sim k\,\exp(-2Ht) \to 0$ is dropped independent of the
original curvature $k=0,\pm 1$.  The CMB causal cone demands $N=\ln
a(t_{\rm end})/a(t_{\rm begin})\geq 60$ (e-folds) between the
beginning and end of inflation. Note that $H(t)$ is determined by the
total potential $V(\phi)=V(0)+\Delta V(\phi)$, i.e., the strength of
the field $\phi$ is an essential input any inflation scenario.}
$\phi(\mu=\LPl) \geq 4.51 \MPl$~\cite{Jegerlehner:2014mua}.  Note that
in the extremely hot Planckian medium, the Hubble constant in the
radiation-dominated state with effective number
$g_*(T)=g_B(T)+\frac78\,g_f(T)=102.75$ of relativistic degrees of
freedom is given by $H=\ell \sqrt{\rho}\simeq 1.66\,\left(k_B
T\right)^2\sqrt{102.75}$ $\mpl^{-1}$. Thus, at Planck time, we have $
H_i \simeq 16.83\,\mpl$, such that a Higgs field of size $\phi_i
\simeq 4.51\, \mpl$, is not unexpectedly large and could as well also
be bigger. This latter requirement looks very conceivable given that
the field is a mode within the hot Planck medium at Planck time. We
know that the Higgs field value is not observable in the LE phase of
the SM. It follows that the value of the Higgs field is the only
unknown parameter of the Higgs potential, and we have to adjust it to
the required amount of inflation.  Herewith, the SM predicts a huge
positive CC at \mbo{\mpl}, given by $$\rho_\phi\simeq V(\phi) \sim
2.77\,\mpl^4\sim \left(1.57\power{19}~\gv\right)^4\epo$$ The CC
supplied by the SM is highly energy-and hence time-dependent, but how
to tame such a huge quasi-constant, which we know is very small today? The
field-dependent part decays exponentially, while the vacuum-energy
$V(0)$ seems to stay large forever. However, as already mentioned
before, here again, the parameters $\lambda$ and $m^2$ are
energy-dependent, causing $m^2 = C_1(\mu)\,\Xi$ to jump to large
negative values at $\mu_0$. Therefore, the CC has to pass a zero as
well. This means we again have a point where the bare and the
renormalized vacuum energy density match. The zero detaches the low
energy region from observable cutoff effects. In particular, any
Planck scale enhancements through quadratic and quartic power
corrections are absent below the zero, where renormalizable QFT is at
work.

\begin{figure}[h!]
\centering
\includegraphics[height=1.6in]{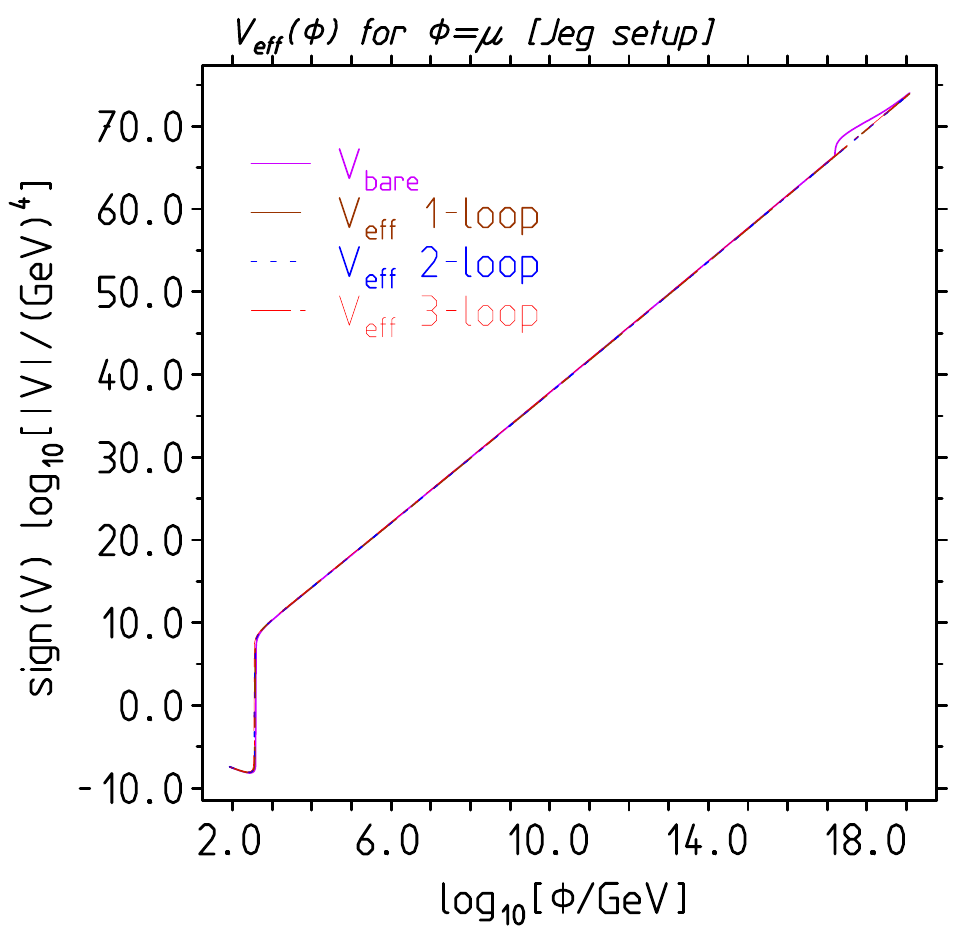}
\includegraphics[height=1.6in]{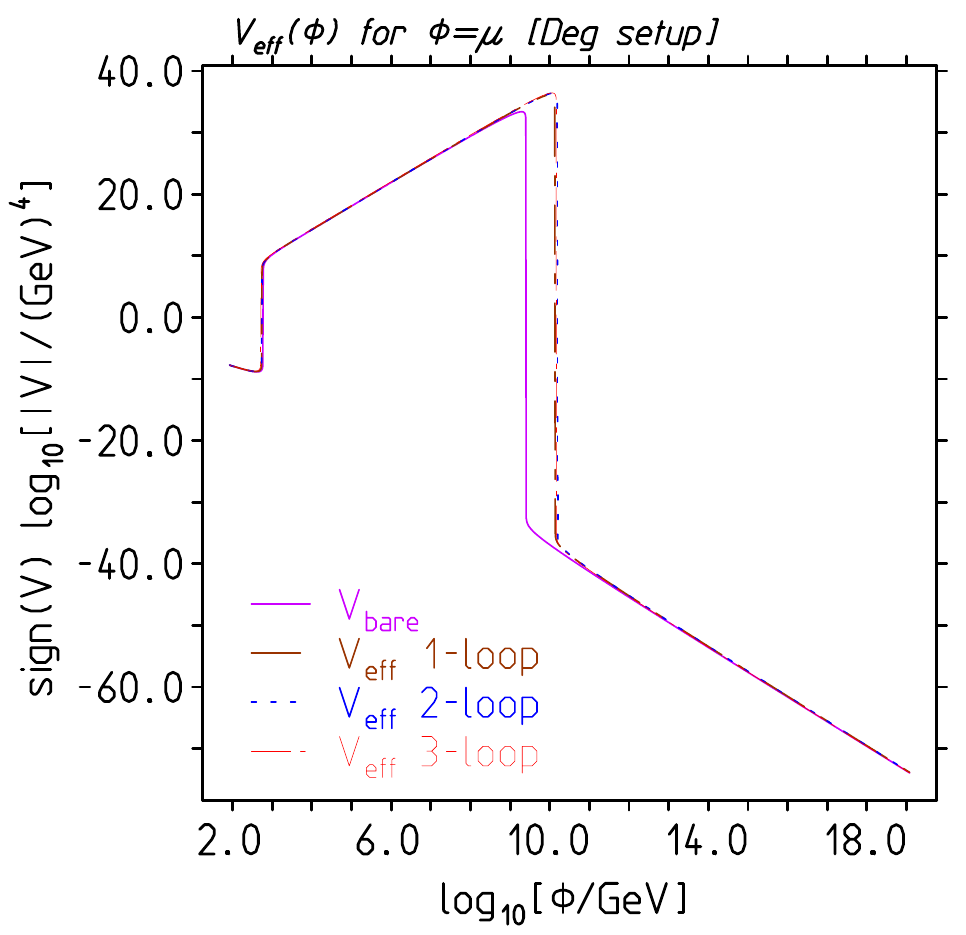}
\caption{The effective potential~\cite{Coleman:1973jx} including
  1-,2-, and leading
  3-loop~\cite{Ford:1992pn,Degrassi:2012ry,Martin:2013gka}
  corrections, with $\mu=\phi$ as a scale. Left: for parameter set
  [{\tt Jeg}] (stable vacuum) in symmetric phase and including Planck-scale enhanced power corrections~\cite{Jegerlehner:2018zxm}. Right: for
  parameter set [{\tt Deg}] (metastable case) in Higgs phase and without power corrections~\cite{Degrassi:2012ry};
  the EW vacuum is tunneling into the bottomless potential.}
\label{fig:VeffLogcomp}
\end{figure}

One more issue we also should touch on. It concerns applying radiative
corrections in the effective potential. Radiative-corrections are
important when Higgs vacuum stability fails, as illustrated in
Fig.~\ref{fig:VeffLogcomp}. In the Low Energy Effective SM (LEESM)
scenario, including the quadratic power correction, and given a stable
vacuum, higher-order corrections have a negligible effect. There is a
big difference in the case of an unstable vacuum where radiative
corrections, in a very narrow parameter interval, yield a positive
extra-contribution to the effective potential, which produces a second
non-trivial minimum of the effective potential, providing a metastable
state. The tunneling time into the metastable state by far exceeds the
age of the universe and hence looks very stable for
us~\cite{Yukawa:3,Degrassi:2012ry}. Nevertheless, an unstable Planck
medium certainly is not a realistic option.

\section{The Hubble constant is highly time dependent}
\begin{figure}[h!]
\centering
\includegraphics[height=2in]{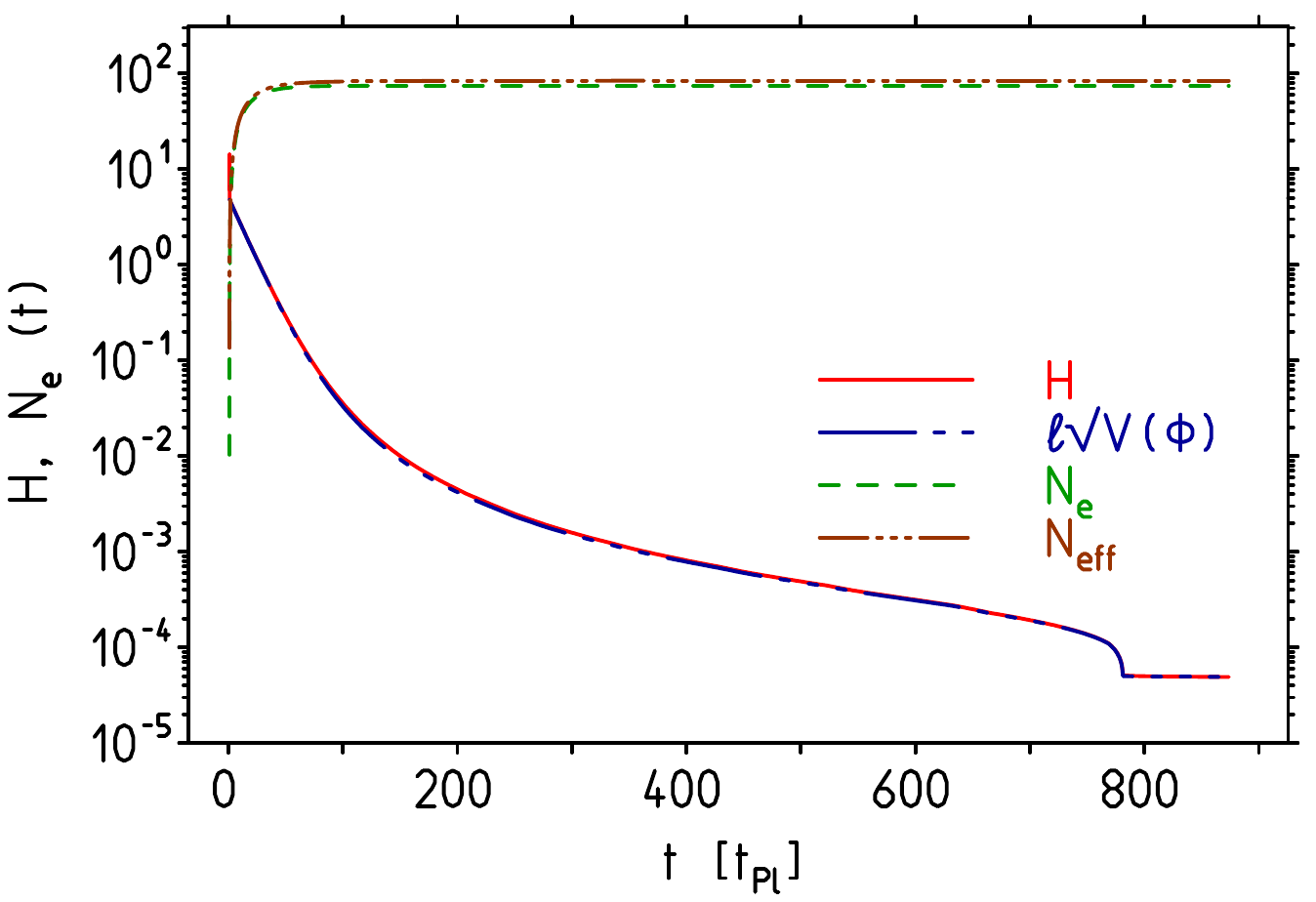}
\caption{The Hubble ``constant'' and the inflation exponent as a
  function of time. $N_e\approx 66$ is reached at time $t\approx
  50\,\tpl\,.$ The Hubble constant $H$ is satisfying $H\approx
  \ell\,\sqrt{V(\phi)}$ very well shortly after Planck time. The
  analytical evaluation of $N_e$ agrees very well with the numerical
  $N_{\rm eff}=\ln a(t)/a(\tpl)$ obtained by solving the coupled set
  of dynamical equations.}
\label{fig:HubbleNeff} 
\end{figure}

As the dark energy density triggering inflation is supplied by the
Higgs system, the equivalent cosmological constant is subject to
dramatic time dependence. The time dependence also substantially
affects the Hubble constant $H$ (see. e.g.~\cite{Poulin:2018cxd} for a
recent discussion) at times when DE has been dominating the total
energy. Then the Hubble constant is dominated by the root of the Higgs
potential $H(t) \approx \ell\,\sqrt{V(\phi)(t)}$, and the
time-dependence is determined by the SM dynamics. Since cold dark
matter is missing in the SM, but it is known to play a prominent role
in shaping our universe, we still have to add it by hand. Yet, the
leading time dependence originates from the highly time-dependent
quadratic and quartic Planck-scale enhancements (see
Fig.~\ref{fig:HubbleNeff}). When the latter vanishes at scale
$\mu_{CC}$, which happens at cosmic time $t_{CC}\approx 2.1
\power{-40}~\mbox{sec}$, also the Hubble constant takes the value we
know from the established Hubble law. The Higgs potential induced
Planck-scale-enhancement effects are missing in the standard
$\Lambda$CDM-model\footnote{The standard Big Bang cosmological model
includes the cosmological constant denoted by $\Lambda$ (representing
the DE), Cold Dark Matter (CDM), ordinary matter, and radiation.} but
certainly play a prominent role in the evolution of the early
universe. It is crucial that the Brout-Englert-Higgs mechanism takes
place long before the time of recombination, when the universe starts
to become transparent and which represents a barrier to direct
observability.  Still, from times before recombination, patterns have
been imprinted to the CMB and the observable matter
distribution. Recently the latter was reviewed more precisely by the
DES-Collaboration~\cite{DES:2022urg}, which may have revealed some
inconsistencies with the $\Lambda$CDM model.

\section{Higgs boson reheating}
In the symmetric phase, except for the Higgs particles, all other
massive SM particles become massless as the Higgs vacuum expectation
value vanishes. The Higgs doublet-field now represents four very heavy
scalars ($H, \varphi, H^\pm$). These states are not only very heavy
with effective mass (see Sect.~\ref{sec:jump} above) $$m(\LPl)\simeq 3.6\power{17}~\gv\,,$$ but
they are also very unstable. They can only decay into fermion pairs
via their Yukawa couplings. Interestingly, the Higgs states cannot
decay into bosons because there are no $HZZ$ and $HWW$ couplings in
the HE phase. Such decays exist only in the LE phase, where the
corresponding couplings are proportional to the Higgs VEV $v$. In the
HE phase, the four Higgs states decay predominantly into those
fermions with the largest Yukawa couplings. It follows that mainly
top-anti-top pairs, but also top-anti-bottom pairs, etc., are
produced, which become very heavy across the electroweak (EW) phase
transition, where they cascade into the lighter fermions according to
the Cabibbo-Kobayashi-Maskawa (CKM) matrix~\cite{CKM}  (see
also~\cite{Glashow:1970gm,KorthalsAltes:1972aq}) and end up in
the baryonic matter. The heavy Higgs particles decay rapidly with
width \ba \Gamma_H \approx \Gamma(H\to t\bar{t}) \simeq
\frac{m}{16 \pi}\,3y_t^2(\LPl) \\ \phantom{\Gamma_H } \simeq 7.4
\power{-3}\,m \simeq 2.7 \power{15}~\gv\,, \ea and lifetime \ba
\tau_H = 1/ \Gamma_H \simeq 2.5 \power{-40} \: \mbox{sec}\,, \ea which
compares to the Planck time $\tpl \simeq 5.4 \power{-44}\:
\mbox{sec}\epo$ The Higgs boson decays reheat the early universe,
which would otherwise have cooled dramatically due to dramatic
inflation.
The maximum matter density in our scenario is
$$\rho_{t{\rm max}} \simeq 1.2 \power{71}~\gv^4 \simeq (0.059\power{19})^4\,,$$
reached at $1.74\,\tpl$, and the reheating temperature is
$$T_{\rm RH} \approx 1.2 \power{-2}\MPl \simeq 2.1 \power{30}~\degK \epo$$
Fortunately, the Higgs bosons cannot
decay into bosons that could disrupt the reheating by what is known as
preheating.  The reader can find details about the reheating in
section 3 of~\cite{Jegerlehner:2014mua}, where it is shown that
Higgs-driven reheating works indeed (for a different scenario,
see~\cite{HiggsReheat18}).

\section{Baryogenesis in the LEESM scenario}
Usually, the violation of baryon number (B) is assumed to occur
due to the pair formation of some unknown heavy particles\footnote{These are often
considered as gauge bosons from a Grand Unified Theory (GUT), where
the SM's gauge structure $G_{\rm SM}\equiv U(1)_Y\otimes \SU(2)_L \otimes \SU(3)_c$
is part of a gauge theory with a simple Lie group $ G_{\rm SM} \subset G_{\rm GUT}$. The simplest possibility is
the group $\SU(5)$, which exhibits 12 additional heavy gauge bosons $X$.} $X$.
One distinguishes two stages: when \mbo{k_B T > m_X} we have thermal
equilibrium, $X$ production, and $X$ decay are in balance; when \mbo{H
  \approx \Gamma_X} and \mbo{k_B T < m_X} $X$-production is
suppressed, and the system is out of equilibrium.

\begin{figure}[h!]
\centering
\includegraphics[height=1.6in]{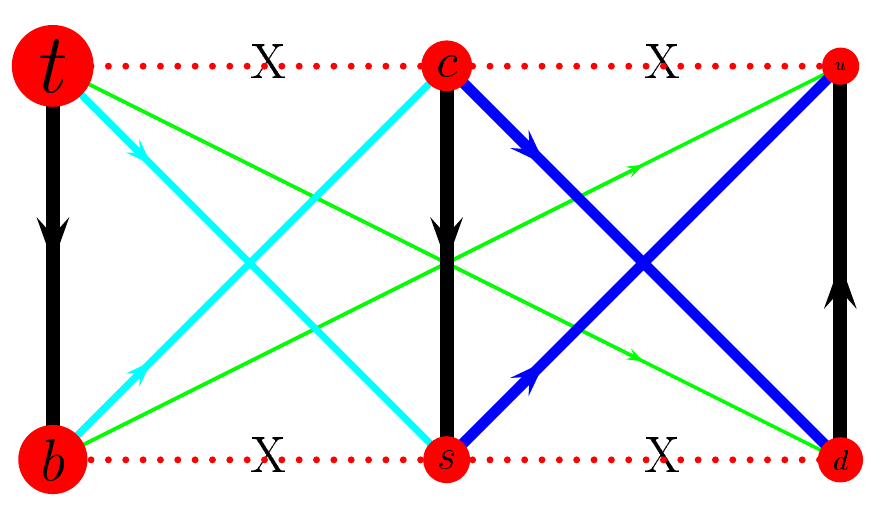}
\caption{ The four heavy Higgs states predominantly decay into heavy
  quarks, which later decay into lighter ones. The top quark decays
  necessarily have to follow the path via the CKM elements
  \mbo{V_{td}} and \mbo{V_{ub}} exhibiting the \underline{leading} CP
  violations. It seems we all are descendants of four heavy Higgs
  bosons and a follow-up intermediate top-bottom quarks-dominated
  gaseous state.}
\label{fig:CKMpat}
\end{figure}

Since the EW phase transition in the LEESM scenario occurs at about
$10^{16}~\gv$ not too far below the Planck scale, higher-order
operators in the low energy expansion come into play. Besides what we
find in the SM, the relevant operators are Weinberg's B-violating
dimension-six four-fermion operators, monomials of the SM
quark-fields~\cite{Weinberg:1979sa}. There are six $B$-violating
operators, which turn out to be $B-L$-preserving. The baryon
matter-antimatter asymmetry, which we need to explain, amounts to
$\eta_B \doteq \frac{n_{X}-n_{\bar{X}}}{n_\gamma }\sim
10^{-10}$.  This effect seems to fit quite naturally into the LEESM
scenario. In the low-energy expansion, the dimension-six operators are
suppressed by \mbo{(E/\LPl)^2}. At the EW phase transition, the Planck
suppression factor is about \mbo{1.3\power{-6}}. Given the
$B$-violation, together with the $C$ and $CP$ violation of the SM, and
the fact that the system during the PT is out of equilibrium, the
conditions for the baryon asymmetry are satisfied~\cite{Sakharov:1967dj}. It appears that the
heavy Higgs bosons are perfect candidates for the required $X$ bosons.
A game-changing difference between the LEESM scenario and the
constellations usually assumed is that baryogenesis for our case
proceeds via the heaviest quarks so that the full CP-violating
``power'' of the CKM scheme (see Fig.~\ref{fig:CKMpat}) is enforced to
come into play.

Besides the CP conserving decays \mbo{H\to
  t\bar{t},\,b\bar{b},\,\cdots} with the predominant Yukawa couplings,
we have the CP violating charged decays \mbo{H^+\to t\bar{d}} [rate
  \mbo{\propto y_ty_d\,V_{td}}] and \mbo{H^-\to b\bar{u}} [rate
  \mbo{\propto y_by_u\,V_{ub}}] which turn out to be important as
well. By the EW phase transition, the interesting decays are \mbo{t\to
  d e^+ \nu} and \mbo{b\to u e^-\nu_e}, etc.

Normal baryonic matter is now a ``trickle-down effect'' from the top
quarks decaying in a cascade into the first-family quarks. Thereby,
CP-violation is most effective not only because the source of normal
matter are the 3rd family quarks but also because it heavily relies on
the fact that Flavor Changing Neutral Currents (FCNC) are forbidden in
the SM~\cite{Glashow:1970gm}. The usual ``unknown'' $X$-particles are
now the known heavy Higgs particles in the symmetric phase of the SM,
excitations of the primordial Planck medium.
All relevant properties of the Higgs ``$X$-particles'' are known: Mass,
width, branching fractions, and CP violation properties! The details
still need to be worked out, but a result of the order of the required
$\eta_B \approx 10^{-10}$ is conceivable. For more details, I refer to
~\cite{Jegerlehner:2014mua}. A convincing investigation remains to be
worked out. However, even in the LEESM framework, $\eta_B$ cannot be
an SM prediction, since the required $B$-violating operators have
unknown couplings. In order to work they should be $O(1)$.  As the
bottom line: baryogenesis (and hence our being) most likely is an ``SM +
dimension six operators'' effect!

\section{There is no naturalness problem with the SM}
At the Higgs phase-transition, when $\delta m_H^2$, and hence the bare
Higgs-potential mass square turns large negative, the vacuum
rearrangement takes place, and all particles (except the photon and
the gluons) acquire their mass. Interestingly, this is true for the
Higgs boson itself.  The Higgs boson mass too turns out to be
proportional to the Higgs field VEV $v$. Note that $v$ stands for
long-range order and becomes zero when $\delta m_H^2$ changes
sign. Since the cutoff effects are nullified at the transition point
$\delta m_H^2=0$, the now renormalizable system has lost its memory of
the cutoff. When there is no connection to $\LPl$, there is no reason
why the Higgs boson cannot be as light as those particles, which are
required to be massless by gauge symmetry and chiral symmetry in the
HE phase. The logarithmic cutoff scale dependence of the
renormalization-group-driven parameters corresponds to a
reparametrization of the renormalizable low energy tail. The
parametrization is renormalization scale-dependent, but a renormalized
QFT is devoid of physical cutoff effects (see Sect.~2
of~\cite{Jegerlehner:2013cta} and references therein).  In place of a
dogma believing that the misunderstood hierarchy
problem\footnote{While the SM has the Higgs VEV $v$ as a
characteristic scale, gravity has its own intrinsic scale $\LPl$. So
one may ask why is $v/\LPl$ so tiny of order $10^{-17}$, but this is
not a problem of the SM. If there is a problem with the SM one has to
ask why the masses and the related couplings are covering 14 orders of
magnitude between the neutrino masses of order $10^{-3}~{\rm eV}$ and
the top-quark mass at $1.72\power{11}~{\rm eV}\epo$ We have no idea
why SM couplings are what they are unless we accept that a conspiracy
is at work, an "observation selection effect" (anthropic principle)
that grants our life in the universe.  The smallness of neutrino
masses, likely, is a consequence of a see-saw mechanism, which could
be triggered within the SM in case the singlets neutrinos would be of
Majorana type. Majorana mass terms are not protected by
symmetry. Therefore, Majorana singlet masses would be Planck-scale
enhanced, similar to the Higgs particle mass in the symmetric phase.}
(see, e.g., ~\cite{'tHooft:1979bh}) is an illness of the SM which must
be cured, e.g., by super-symmetrization of the SM as one possibility,
Veltman has stressed many times the fact that within the minimal SM,
the zero photon mass is a prediction, not subject to renormalization,
while in most SM extensions this prediction is
lost~\cite{DiazCruz:1992uw} and has to be replaced by an extra
condition\footnote{The simplest supersymmetric model accidentally
escapes this problem and predicts a zero photon mass as well.}, i.e.,
it has to be fine-tuned~\cite{Veltman:2008zz}.

Indeed, most extensions of the SM require non-minimal Higgs
sectors\footnote{It also tells us that renormalizability alone, by
far, does not make a model to be naturally emergent. Also, symmetries
by no means always are emergent. The most prominent example is
$U(1)_{\rm QED}$, which is unavoidably emergent only in a minimal
Higgs scenario, i.e., in conjunction with the weak- and Higgs-sectors,
which reveals the hidden tree-level custodial symmetry. The latter is
responsible for another specific SM feature, namely the tree-level
relation $\cosW\,\mz/\mw=1$, between the weak gauge boson masses
$M_W$, $M_Z$ and the weak mixing parameter $\sinW=1-\cosW$, which gets violated by
most SM extensions and then creates a different fine-tuning
problem~\cite{Czakon:1999ha}.  Another example is the $R$-parity
usually imposed in SUSY models to prevent FCNCs and to enforce the
existence of a CDM candidate, $R$-parity is not something a LEE
produces for free.}: Two Doublet Higgs Models (TDHM), SUSY extensions
like the Minimal Supersymmetric SM (MSSM), Grand Unified Theories
(GUT), left-right symmetric models, the
Peccei-Quinn~\cite{Peccei:1977hh,Preskill:1982cy} approach to the
strong CP problem, inflaton models base on an extra scalar field and
many more. The masslessness of the photon is one of the most basic
facts of life (no life otherwise), so why should we make this
fundamental property to be something we have to arrange by hand?

Veltman in~\cite{Veltbrighton} has shown that in the SM with the
minimal Higgs sector violations of parity $P$ and charge conjugation
$C$ as they appear in the SM are emergent (automatic)
properties. Given three families of fermions, the violation of $CP$ is
also emergent. This is a strong indication that these discrete
symmetries cannot be violated within the Planck medium. Otherwise, the
symmetry-breaking pattern realized in the SM would differ from what
the minimal SM predicts. A $P$, $C$, and $T$ invariant primordial
medium, however, is able to explain the absence of the $P$, $C$, and
$CP$ violations in the strong interactions. Therefore, there is no
$\theta$-term in the QCD-Lagrangian, since such a term is not
emergent, i.e., it does not arise automatically as a relict in the
low-energy expansion. A lattice QCD simulation finds that a
non-vanishing $\theta$ term would spoil confinement~\cite{Nakamura:2021meh},
and therefore must be absent.

\section{The QFT and gauge field structure of the SM are a natural consequence of the LEET scenario}
The strong indication that the SM is a low energy effective relict of
a cutoff system at the Planck scale has a dramatic impact insofar that
almost all the ``\textit{axioms}'' (fundamental principles) which
define the renormalizable QFT structure are consequences of the
blindness for the details inherent in the largely unknown Planck
medium. Here universality is at work. The LEET is universal for a
broad class of systems of different short-distance properties. After
all, in the low-energy expansion, one loses all information carried by
the infinite tower of the positive powers of $x=E/\LPl$. What remains
observable we know to be:\\

\myitem{{\bf 1)}} Local quantum field theory emerges as a universal
long-range structure of cutoff media like classical spin systems (the
ingenious discovery of Ken Wilson~\cite{Wilson:1971bg}). Euclidean space
and Minkowski spacetime field theories share the property of
analyticity and hence are equivalent (Osterwalder-Schrader theorem),
connected by a Wick
rotation~\cite{Wick:1954eu,Osterwalder:1973dx}. Therefore, what we
experience as \textit{time}, likely is an emergent phenomenon as well.

\myitem{{\bf 2)}} Low energy effective remnant interactions exhibit local
  non-Abelian gauge theory
  structure~\cite{Cornwall:1973tb,LlewellynSmith:1973ey,Bell:1973ex,Bass:2017nml}.
  The non-Abelian gauge symmetries appear as a result of so-called
  gauge cancellations, which eliminate effects exhibiting bad
  (non-renormalizable) high-energy behavior.

Here, one big puzzle opens up: how is it possible to have a strictly
massless photon? The Abelian $U(1)_{\rm QED}$ cannot be emergent like
the non-Abelian counterparts. For the Abelian case, there are no gauge
cancellations. Even a QED with a massive photon (massive neutral
vector meson theory) remains
renormalizable~\cite{Stueckelberg:1957zz,Symanzik68}. Hence,
renormalizability alone cannot explain the strictly massless photon.
The answer is that, for the minimal Higgs SM, a massless photon is
unavoidable~\cite{Veltbrighton} because, in any case, the spin-1
mass-matrix exhibits a zero eigenvalue, protected by Abelian gauge
invariance. The latter is a residual symmetry within its context.

\myitem{{\bf 3)}} The LEET is renormalizable and non-trivial, and all
properties required by renormalizability are consequences. The minimal
renormalizable extension of ``QED plus Fermi weak theory'' fixes the
SM with all its stunning properties quite
unambiguously~\cite{Glashow61,Weinberg67,Glashow:1970gm,KorthalsAltes:1972aq,QCD}.
While one fermion family seems sufficient to guarantee our daily
existence, two more replicas are mandatory because we need CP
violation for the baryogenesis~\cite{Sakharov:1967dj,CKM}. The third
family also hosts the heavy top quark, which must exist to cooperate
with the heavy gauge bosons and the Higgs-boson, as we have seen
above. Note that non-trivial (i.e., interacting) relicts of a
low-energy expansion only appear in spacetime-dimension $D=4$ or
below.  Non-triviality excludes spacetime dimensions $D>4$ from being
observable; only hidden non-interacting free fields would exist
there.\\

Local gauge symmetry requests particles to conspire in multiplets, and
the most likely natural choice is doublets, triplets separate from
singlets, as we have them in the SM. Renormalizability requires the
existence of at least one scalar field, i.e., as a minimal choice, we
precisely get the Higgs boson as we know it from the
SM. Veltman~\cite{Veltbrighton} has presented a convincing derivation
of the amazingly looking SM structure from general properties we find
unveiled as automatic in a minimal LEET scenario. A more elaborate
discussion the reader may find
in~\cite{Jegerlehner:2018zxm,Jegerlehner:2021vqz}.

Since the LEESM scenario is supported seriously by many
phenomenological facts, attempts to quantize gravity the same way as
the other three interactions look obsolete because relativistic QFT is
a consequence of the low-energy expansion, and so is quantum mechanics
(QM) as the non-relativistic limit of QFT (see
also~\cite{tHooft:2014znz,tHooft:2022pra} for a different lattice-system approach
to QM). They are emergent properties far below $\MPl$. The Planck
medium itself is a cutoff system, and there is no reason to believe it
should look like a QFT or an extension like string theory. In
principle, at low energies, gravity would show up as spin-2
graviton\footnote{ These would be a natural product of a multipole
expansion of a potential provided by the Planck
medium~\cite{Jegerlehner:1978nk}.}, but, as we know, there is no
renormalizable theory of spin-2 fields. Therefore possible graviton
effects are beyond being observable.

\section{SM Higgs-inflation confronts physics beyond the SM}
The SM has no answer for the existence of cold dark matter, but the
latter could result from the underlying Planck cutoff
medium. Similarly, effective interactions outside the SM (as sketched
above) must supply baryon number violation. Besides these evident
``deficiencies'', the LEESM scenario, with the Higgs field
representing the inflaton that provides sufficient dark energy,
appears convincingly supported by the data-driven bottom-up
approach. However, the scenario only works if the SM conspiracy is not
disturbed by such beyond the SM effects that significantly would
change the delicate picture of the running parameters and
significantly reduce the DE supply in the symmetric
phase. Consequently, all SM extensions figured out to solve the
presumed SM hierarchy- or naturalness problem, must be given
up. Corresponding models have been invented to kill or substantially
reduce the quadratic and quartic power enhancements
\footnote{Usually identified as quadratic and quartic UV
singularities. In our case, the cutoff is physical and finite,
although very large.}, which in the LEESM scenario promote the Higgs
to be the inflaton. Even a fourth family of fermions would spoil Higgs
inflation because a fourth family would exhibit a heavy quark $t’$ and
a corresponding Yukawa coupling larger than $y_t$ and demolish the
pattern of the SM running parameters. At least, this is what
experimental bound on heavy fermions suggest strongly. Such extensions
make early cosmology a matter of unknown physics beyond the SM. So,
one usually ad hoc adds an inflaton field with unknown properties,
which then are tuned to fit observation. In contrast to structures
like string theory or quantum gravity that are beyond being testable,
the LEESM setting is easy to disprove: find supersymmetric (SUSY),
grand unification (GUT), two-doublet Higgs (TDHM) extensions of the SM
or a fourth-family fermion, and the whole picture collapses!  Also,
the predictions following via the SM extrapolations concerning
inflation and other cosmology issues can falsify the model if they
fail to fit observations. The high sensitivity implies a high
predictive power, although it appears highly fine-tuned, which is no
contradiction within the emergence paradigm. We have a chaotic hot
system at the Planck scale, and obviously, only what can pervade long
distances arrives there, namely a spectrum of comparably light
particles. The filtering process acts to let it appear as a
self-organized system~\cite{wiki}. I think that the LEET approach and
Veltman's explanation of the SM structure provide convincing support
to understand why it is the SM that passes the low-pass filter, which
obscures the details of the medium it comes from.

\section{Epilogue, quo vadis?}
It seems commonly accepted that the SM Higgs vacuum is unstable with
intermediate
metastability~\cite{Yukawa:3,Degrassi:2012ry,Bednyakov:2015sca,Kniehl:2015nwa}.
Then the Higgs boson within the "General Relativity plus SM" framework
cannot be the inflaton\footnote{A very different model of Higgs
inflation, which has barely something in common with our LEESM
scenario, is the Minkowski-Zee-Shaposhnikov et
al.~\cite{Minkowski:1977aj,Zee:1978wi,Bezrukov:2007ep,Barbon:2009ya,Bezrukov:2010jz,Bezrukov:2014bra,Bezrukov:2014ipa}
so-called \textit{non-minimal SM inflation} scenario. It is based on a
modification of Einstein's theory of gravity (see
also~\cite{Hamada:2015ria}).  A time-dependent cosmological constant
has been considered also in a model which is based on a dilatation
symmetry anomaly, where one assumes the Newton ``constant'' to be a
time-dependent dynamical degree of
freedom~\cite{quintessence,Wetterich:2014gaa}.}.  However, why should
the precise value of the Higgs mass\footnote{At which value of the
Higgs mass vacuum stability holds depends on the other relevant SM
parameters, and in particular on the top-quark mass, which is not
simple to extract from the top-quark production
data~\cite{Beneke:2016cbu}.} fail to provide ``vacuum stability up to
$\LPl$'', and this by 1.5 $\sigma$ only, as some analyses
report\footnote{Usually, evidence only is granted when a 5 $\sigma$
gap gets established.  A much lower ``discrepancy'' is a problem,
especially when there is a possible dramatic instability nearby, as in
our case.}? One should be aware that, very likely, the uncertainties of SM
parameters have been underestimated, as some known inconsistencies
persist.  In fact, some ``established'' SM parameters reveal
differences at a 2 $\sigma$ level, and one has sufficient flexibility
allowing the possibility outlined in my LEESM outcome. The SM as an
attachment of a Planck medium is highly supported by the data-driven
evidence-based approach. There is a good chance that vacuum stability
will be confirmed in the future, achieved by progress in determining
more precise input parameters and higher-order calculations. It would
mean that the Higgs discovery is a true game-changer opening the door
for SM-driven early cosmology.

Keep in mind: the Higgs mass miraculously turns out to have a value
that has been expected from vacuum stability arguments. It looks like
a tricky self-organized conspiracy between the relevant SM couplings
to reach this purpose. And if it would miss stabilizing the vacuum,
why then the gap from the stable vacuum is so minor?

The big issue is the very delicate \textit{conspiracy between SM
  couplings}: the precision determination of parameters is more
important than ever. A big challenge for LHC and ILC/FCC (see,
e.g.,~\cite{Accomando:1997wt,Proceedings:2019vxr}) are the precision
values for \mbo{\lambda}, \mbo{y_t} and \mbo{\alpha_s}. A future
Higgs-boson/top-quark factory will be required to grant the necessary
goals here. Also, the non-perturbative QCD sector of the SM remains a
big challenge. The precision of the calculations of the electroweak
running couplings \mbo{\alpha(E)} and \mbo{\alpha_2(E)} sensibly
depend on the proper inclusion of strong interaction effects. More
precise hadronic cross-section measurements at low-energy hadron
facilities are needed to reduce hadronic uncertainties, which are
limiting the precision of the standard data-based dispersion relations
evaluations. Soon, \mbo{\alpha(E)} and \mbo{\alpha_2(E)} are going to
be determined with improved accuracy reliably by lattice QCD (LQCD)
calculations (see~\cite{SanJosePerez:2022qwc} and references
therein). The advantage of LQCD ab-initio calculations is that only
this method allows for unambiguously separating the pure QCD effects
from QED effects, which are mixed up in experimental cross-section
data and are not easy to disentangle.

\noindent
Outlook:\\ Since the Higgs particle (and its field) is the only known
particle that directly couples to gravity via the energy-momentum
tensor of the Einstein-equations, the SM Higgs sector may mediate a
direct gate to the precision cosmology of the early universe! But this
option requires the SM to be extendable up to the Planck scale and
that the Brout-Englert-Higgs mechanism got triggered below the Planck
scale, i.e., the SM has been in the symmetric phase there. Only the
discovery of the 125 GeV Higgs boson has opened this door for
Higgs-cosmology. In any case, -- ``the SM Higgs boson in the Planck
window'' -- provides strong support that the SM is a naturally
emergent structure and that the extrapolation of the SM up to the
Planck scale could reveal the Higgs boson as \textit{the master of the
  universe}.

\noindent
Acknowledgments:\\ I thank Oliver Bär for his critical review of the
manuscript and for his useful suggestions and questions.


\end{document}